\documentclass[sigplan,nonacm]{acmart}
\AtBeginDocument{%
  }

\usepackage{xspace}
\usepackage{pifont}
\usepackage{array}
\usepackage{tabularx}
\usepackage{subcaption}
\usepackage{amsthm}
\usepackage{etoolbox}
\usepackage{listings}
\usepackage{multirow}
\usepackage{makecell}
\usepackage{enumitem}
\usepackage[most]{tcolorbox}
\usepackage{balance}

\setlist{nolistsep}

\newcommand{\para}[1]{\smallskip\noindent\textbf{#1}}

\def\Snospace~{\S{}}

\newcommand{\cmark}{\ding{51}}%
\newcommand{\xmark}{\ding{55}}%

\newcommand{\sys}{Ossim\xspace}

\newcommand{\sysbin}{ossim}
\newcommand{\syscmd}{\texttt{\sysbin}\xspace}
\newcommand{\sysdc}{\texttt{\sysbin d-\allowbreak controller}\xspace}
\newcommand{\sysdw}{\texttt{\sysbin d-\allowbreak worker}\xspace}
\newcommand{\syshostctl}{\texttt{\sysbin-\allowbreak hostctl}\xspace}
\newcommand{\sysagent}{\texttt{\sysbin-\allowbreak guest-\allowbreak agent}\xspace}

\newcommand{\syslib}{\texttt{lib\sysbin}\xspace}

\definecolor{googleRed}{HTML}{EA4335}
\definecolor{googleGreen}{HTML}{34A853}
\definecolor{googleBlue}{HTML}{4285F4}

\newcommand{\cmt}[1]{}

\newif\ifshowcomments

\ifshowcomments
  \newcommand{\editorial}[2]{{\textcolor{#1}{#2\xspace}}}
  \newcommand{\todo}[1]{\editorial{googleRed}{[TODO: #1]}}
  \newcommand{\budget}[1]{\texorpdfstring{{\normalfont\small\textcolor{magenta}{~[budget: #1]}}}{}}
  \newenvironment{new}{\color{green!70!black}}{}
  \newenvironment{old}{\color{black!30!white}}{}
\else
  \newcommand{\editorial}[2]{}
  \newcommand{\todo}[1]{}
  \newcommand{\budget}[1]{}

\fi

\begin{document}

\title[]
{\sys: OS-Driven Live Simulation for Cluster-Scale Full-Stack Evaluation}

\author{Yiliang Wan\textsuperscript{1},
Jonas Kaufmann\textsuperscript{2},
Yihan Yang\textsuperscript{1},
Chaoyi Ruan\textsuperscript{1}, \\
Haifeng Sun\textsuperscript{1*},
Antoine Kaufmann\textsuperscript{2},
Jialin Li\textsuperscript{1}}
\affiliation{%
  \vspace{8pt}
  \institution{\textsuperscript{1}National University of Singapore \qquad \textsuperscript{2}Max Planck Institute for Software Systems}
  \country{ }
}
\renewcommand{\authors}{Yiliang Wan, Jonas Kaufmann, Yihan Yang, Chaoyi Ruan, Haifeng Sun, Antoine Kaufmann, Jialin Li}
\makeatletter
\g@addto@macro\@authornotes{\stepcounter{footnote}\footnotetext{Corresponding author.}}
\makeatother

\renewcommand{\shortauthors}{Wan et al.}

\begin{abstract}

Cluster-scale full-stack simulation is essential for evaluating distributed software stacks and emerging hardware components before deployment.
Such simulation must achieve both full-stack fidelity for the unmodified production stack and the simulation performance required for iterative configuration exploration.
However, no existing method achieves both.
We present \sys, an OS-level approach to cluster-scale full-stack simulation built on top of the Linux virtualization stack.
\sys comprises four subsystems: simulation-oriented scheduling, 
live memory hierarchy management, simulation-aware IPC,
and multi-host coordination.
Together, they coordinate live and modeled components under shared simulated time while controlling interference among co-located live VMs.
These mechanisms point toward \emph{simulation-native OS support}, where simulation control and orchestration become core OS responsibilities.

\end{abstract}

\maketitle

\section{Introduction \budget{1}}

Evaluating a distributed system requires understanding interactions across its stack: applications and runtimes, operating systems, processors and memory hierarchies, devices, and the network.
These interactions determine whether a deployment configuration improves application performance and whether an emerging hardware capability delivers its expected benefits.
Cluster-scale full-stack simulation enables such evaluation before hardware~\cite{tpu,accelnet,accelflow,cxl-real-devices,cxl,ualink,uec} is available at the target scale or production deployment is practical.
It also facilitates choosing deployment configurations for production workloads such as cloud platforms~\cite{openstack,k8s} and big-data frameworks~\cite{spark,hadoop}.
To support iterative exploration, however, a simulator must both preserve the behavior of the production stack and execute at practical speed.

Meeting these requirements presents two challenges.
The first is hardware feature coverage.
Production software can depend on specific processor instructions and features, architectural mechanisms, or device capabilities such as NIC hardware offloading.
Faithfully evaluating such software requires the simulator to implement and precisely model the needed hardware features or functionalities.
Existing simulators do not necessarily support the required features, even when they provide detailed models of other aspects of the system.
Disabling an unsupported feature or substituting a simplified implementation can change the workload. 

The second challenge is the cost of detailed host simulation.
Full-system simulators can execute unmodified operating systems, and modular frameworks such as SimBricks~\cite{simbricks} and SplitSim~\cite{splitsim} compose host, device, and network simulators into end-to-end systems.
However, detailed CPU and memory-system simulation can dominate execution time and constrain the progress of the entire composition.
Accurately modeling host performance requires capturing not only instruction execution but also cache misses, coherence traffic, memory contention, and their overlap with computation.
Resolving these interactions in software is expensive.
Simplifying them improves simulation speed but can distort application execution rates and, consequently, the load and timing observed by other components.

Live simulation offers a complementary approach that addresses both challenges.
Live simulation executes available components directly on physical hardware.
For features exposed through the live execution interface, it reuses the underlying hardware instead of requiring a simulator to implement that functionality.
For processors and memory, native execution lets the physical hardware perform instruction execution, cache lookup, coherence, and memory access directly, avoiding the corresponding fine-grained simulation work.
When the available hardware is representative of the host being studied, this preserves its native behavior while reserving explicit models for components or properties that differ from the target system.

However, native execution alone is not simulation.
A VM's progress depends on host scheduling and resource availability, while a modeled component advances according to its own timing model.
Uncontrolled differences between these time bases can alter guest timers, communication timing, and distributed coordination.
Likewise, using physical memory preserves the host's memory behavior but does not automatically reproduce a different target latency.
Combining live and modeled components therefore requires coordinated control over execution progress, resource performance, and the simulated times at which interactions become visible.

The key challenge is making virtualized hosts and heterogeneous modeled components composable under coordinated simulation control.
Our focus is therefore a \emph{composable substrate} in which VMs and modeled components share execution and communication abstractions, live resources can incorporate performance adjustments, and extending the mechanisms across physical hosts.
These controls must work together: changing a vCPU's effective execution rate must remain consistent with its guest-visible clock, its scheduling relative to other components, and its external interactions.

Our key insight is that the operating system with hardware virtualization support~\cite{kvm,qemu,libvirt} is the natural integration point for these controls.
The host OS encompasses both application and guest-kernel execution, allowing an unmodified software stack to participate in simulated time.
Hardware virtualization retains native execution and access to exposed processor features, while the host controls vCPU scheduling, guest-visible clocks, and the resources backing guest memory.
Virtual-device interfaces provide additional integration points for modeled communication and external components.
Coordinating these mechanisms allows us to reuse native hardware behavior while controlling the progress and interactions of complete virtualized hosts.

We present \sys, an OS-native substrate for full-stack live simulation at cluster scale. \sys builds on the bases of prior live simulation systems~\cite{diecast,slicetime,timekeeper}, and follows two design principles.
First, it integrates simulation control into OS mechanisms, preserving efficient native execution while enforcing simulated-time progress, resource-performance adjustments, and timed event delivery.
Second, it defines common abstractions and interfaces that separate generic simulation control from component-specific implementations and performance models.
This separation enables flexible composition across live and modeled components, host-backed resources, and communication models without redesigning the underlying control mechanisms.
Our design comprises four major subsystems:

\begin{itemize}[leftmargin=*]
    \item \emph{Simulation-oriented scheduling} represents live vCPUs and modeled components through a common virtual-time abstraction.
    Skew-bounded scheduling coordinates their progress, while gang scheduling supports live components that require coordinated execution.

    \item \emph{Live memory hierarchy management} retains native memory accesses and coherence, manages their backing resources, and uses hardware-counter measurements to adjust virtual-time advancement for configured memory-latency differences.

    \item \emph{Simulation-aware communication} provides common interfaces that connect live guests with built-in models or external simulators, and controls when messages become visible to receivers.

    \item \emph{Multi-host coordination} extends progress tracking and timed communication across machines, allowing a simulation to use hardware resources beyond a single host.
\end{itemize}

We implement \sys in the Linux virtualization stack and evaluate it with unmodified distributed services, databases, and analytics frameworks.
Across seven reported end-to-end metrics, \sys reproduces corresponding physical-system measurements within 9\%, while incurring only 2.6-24$\times$ wall-clock slowdown.
In separate simulation-performance experiments, \sys incurs 5.5-25$\times$ slowdown, compared with 665-1426$\times$ for a gem5 TimingSimple-based modular configuration and 3381-9329$\times$ for an O3-based configuration.
These results demonstrate that coordinating native execution, live resources, and modeled interactions through OS support can make full-stack simulation practical without requiring detailed software simulation of the entire host.

\section{Background and Motivation\budget{1.25}}
\label{sec:background}

\subsection{Cluster-Scale Full-Stack Simulation}
\label{sec:background:requirement}

Cluster-scale full-stack simulation enables evaluating distributed software stacks and hardware configurations before the target system is available at scale or production deployment is practical.
Such evaluation must capture both the functionality of individual components and their cross-layer interactions.
For example, offloading network processing to a DPU changes not only packet-processing latency but also the CPU resources available to a distributed database.
Evaluating its end-to-end benefit therefore requires capturing the production software stack together with processor, memory, device, and network behavior.
A practical simulator must preserve these relevant behaviors while running fast enough to support iterative exploration.
Therefore, we have two requirements for a cluster-scale full-stack simulator:

\para{R1. Full-stack fidelity.}
We distinguish two complementary aspects of fidelity: full-stack coverage and accuracy.

\begin{itemize}[leftmargin=*]
\item
\emph{Full-stack coverage} concerns the functionalities and interactions the simulation represents.
On the software side, it requires executing the entire stack without source modification;
on the hardware side, coverage requires the processor instructions, architectural mechanisms, and device capabilities exercised by that software.
Replacing an end host with a traffic generator or disabling an unsupported hardware feature can change the software paths and cross-layer behavior being evaluated.

\item
\emph{Accuracy} concerns how faithfully the represented system behaves relative to the target.
Execution rates, resource contention, and communication timing must reflect the target closely enough to produce representative end-to-end results.
For example, exposing an offload interface without capturing its effects on host CPU demand and packet-processing time would produce inaccurate results.
\end{itemize}

\para{R2. Simulation performance.}
The simulator must complete end-to-end experiments fast enough for iterative configuration exploration.
Users evaluate a candidate hardware, network, or software configuration, inspect its behavior, and refine it across repeated experiments.
At cluster scale, each experiment executes a complete production stack on every simulated server, so the wall-clock time required for the overall simulation becomes critical: if a minute of target execution requires days to simulate, exploring multiple configurations becomes impractical.

\subsection{Related Simulation Methods}
\label{sec:background:existing}

\para{Discrete event simulation (DES).}
DES models the target system as a sequence of events processed in simulated time.
Host simulators such as gem5~\cite{gem5} and Simics~\cite{simics} execute complete software stacks with models of processors, memory systems, and devices, while network simulators such as ns-3~\cite{ns3} and OMNeT++~\cite{omnetpp} model protocols and topologies.
These tools provide detailed behavior within their supported configurations.
However, a layer-specific simulator alone does not capture the interactions across an entire distributed system.
Network simulations, for example, often use traffic generators or simplified hosts instead of executing the production software stack.

DES also presents two practical challenges.
First, functional coverage depends on the simulator implementation, which may omit certain hardware instructions or mechanisms.
Second, detailed timing models can be expensive to execute.
For example, simulating one second of a 1,024-server fat-tree DCTCP workload in ns-3 can take over ten hours~\cite{unison}.

\para{Modular composition.}
Modular frameworks~\cite{simbricks,splitsim} connect independently developed component simulators through common interfaces and simulated-time synchronization.
They enable host, device, and network models to form an end-to-end system while retaining their implementations.

Composition does not resolve limitations within those components.
Required hardware functionality must still be supplied by a simulator, and the overall speed is determined by the slowest component simulator, which is typically end-host simulation.
Modularity itself does not inherently remove the cost of detailed CPU and memory-system modeling.

\para{Live simulation.}
Live simulation executes physically available components directly on hardware.
This allows live simulation to reuse the hardware features, removing the dependency on software implementations.
Native processor execution also lets hardware perform instruction execution, cache lookup, coherence, and memory access directly;
this preserves the native behavior of processors without fine-grained host simulations.
Explicit models are then only required for components that are unavailable.

However, live components execute according to physical scheduling and resource availability, whereas modeled components advance based on their timing models.
Moreover, the available hardware may differ from the target in resource performance.
To ensure accurate simulation, live simulation must control execution progress, account for resource differences, and coordinate when interactions become visible.

Prior systems establish key foundations for this approach.
VM-based systems~\cite{diecast,slicetime} demonstrate execution of unmodified software stacks under controlled time, but with simulation control tailored only to particular settings.
Process- and container-based systems~\cite{timekeeper,timerkeeper-intg,phantora} provide virtual-time control for native applications, but leaves the guest OS and interactions with hardware outside the simulated stack.
Other systems~\cite{nex,phantora} explore live execution for accelerated workloads and hardware/software evaluation, but only target specialized stacks.

\subsection{Toward a General Live-Simulation Substrate}
\label{sec:background:opportunity}

Modular composition and live execution suggest a general \emph{live-simulation substrate}, which reuses hardware for native functionality and efficient execution, while exposing common interfaces for composing heterogeneous components.
The key challenge is that the substrate must coordinate execution and interactions among live and heterogeneous modeled components in a unified manner, while retaining the compatibility and performance advantages of native execution.

Realizing this substrate requires four capabilities.
\begin{itemize}[leftmargin=*]
    \item \emph{C1: Simulated-time execution control} must coordinate live and modeled execution according to simulated rather than wall-clock time, with a common representation of simulated progress across different execution types.
    \item \emph{C2: Live-resource control} must prevent unintended host interference and adjust the performance observed from host-backed resources when it differs from the target system.
    \item \emph{C3: Timed cross-component interaction} must provide common communication interfaces and control the simulated times at which interactions between components become visible.
    \item \emph{C4: Multi-host coordination} must extend simulated-time progress and timed communication across physical hosts without changing the abstractions exposed to the simulation.
\end{itemize}

These capabilities are interdependent.
Adjusting a live processor's execution rate to account for target memory latency must also affect its guest-visible time, its progress relative to other components, and the timing of its external interactions.
Likewise, a pending message must remain coordinated with the receiver's progress even when it crosses a physical host boundary.
The central challenge is therefore to integrate execution, resource, and interaction control while keeping their common mechanisms independent of component-specific implementations and performance models.

\section{\sys Overview\budget{1}}
\label{sec:overview}

\subsection{OS as a General Live-Simulation Substrate}
\label{sec:overview:os}

\begin{figure}[tp]
\centering
\includegraphics[width=.8\linewidth]{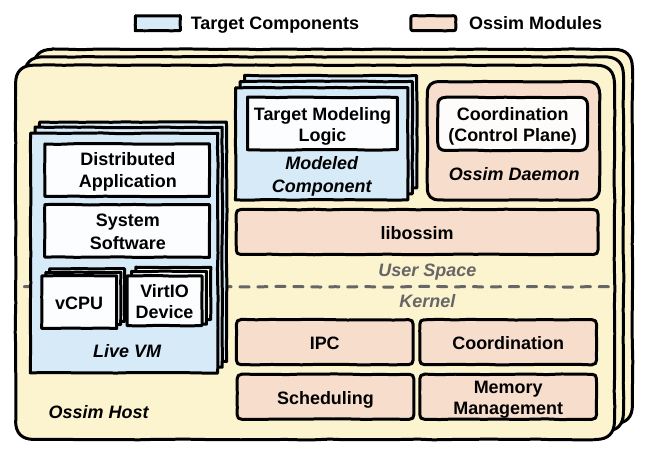}
\vspace{-5pt}
\caption{The overall architecture of \sys.}
\vspace{-10pt}
\label{fig:architecture}
\end{figure}

Our key observation is that the host OS is the natural integration point for all four
capabilities identified in \autoref{sec:background:opportunity}.
It provides common control over the execution, physical resources, and communication
of heterogeneous live and modeled components, making it possible to provide
general simulation abstractions independent of specific components. 

\para{Simulated-time execution control.}
Live and modeled components must advance under a common notion of simulated progress.
Live VMs execute through a hypervisor under host-OS control,
while modeled components typically run as user-space processes scheduled by the kernel.
The OS therefore provides a common point for coordinating their execution and time.

\para{Live-resource control.}
Live components directly use physical resources such as CPUs, memory, caches, and devices.
Since the OS manages their allocation, isolation, and access, it is the natural point
for controlling host interference and the resource behavior exposed to simulation.

\para{Timed cross-component interaction.}
Interactions between components must become visible according to simulated time.
Although components may use different communication mechanisms, their interactions rely
on OS-managed communication and virtual interfaces.
The host OS thus provides a common point for controlling interaction timing across heterogeneous components.

\para{Multi-host coordination.}
At cluster scale, the same execution and interaction semantics must hold across physical hosts.
Because these semantics are enforced at each host OS, extending them across machines
naturally requires coordination among the participating OS instances.

Together, these observations motivate the host OS as the foundation for a general
live-simulation substrate: it provides common control points spanning heterogeneous
components, resources, interactions, and physical hosts.

\subsection{\sys Architecture}
\label{sec:overview:architecture}

\autoref{fig:architecture} shows the architecture of \sys.
We implement the OS-native substrate described above in the Linux virtualization
stack~\cite{qemu,kvm}.
Within this stack, hardware virtualization~\cite{intel64,amd64} retains native execution
and access to exposed processor features, while the host OS controls vCPU scheduling,
guest-visible clocks, and the resources backing guest memory.
Virtual-device interfaces further connect live VMs with modeled components.
Building on these control points, \sys integrates simulation control into the OS 
while separating generic mechanisms from component-specific implementations.

\sys comprises four subsystems.
\emph{Simulation-oriented scheduling} represents live and modeled components
through a common virtual-time abstraction, with domain-based scheduling
coordinating their progress (\autoref{sec:scheduling}).
\emph{Live memory hierarchy management} retains native memory accesses and coherence,
manages their backing resources, and adjusts virtual-time advancement for target
memory-performance differences (\autoref{sec:memory}).
\emph{Simulation-aware IPC} provides common interfaces connecting live guests with
heterogeneous communication fabric implementations, 
while controlling when messages become visible to receivers (\autoref{sec:ipc}).
Finally, \emph{multi-host coordination} extends the above mechanisms
across physical machines, allowing a simulation to use resources beyond a single host
(\autoref{sec:multihost}).

Together, these subsystems realize a general live-simulation substrate that preserves
native execution where possible,
 while bringing execution, resources, and interactions under shared simulation control.
\autoref{tab:comparison} compares \sys with existing methods.

\begin{table}[t]
\centering
\caption{Comparison of simulation methods for cluster-scale full-stack simulation.}
\label{tab:comparison}
\vspace{-5pt}
\footnotesize
\setlength{\tabcolsep}{2.5pt}
\begin{tabular*}{\columnwidth}{@{\extracolsep{\fill}}llccc@{}}
\hline
Method        & System             & Sim.        & End-host   & Cluster- \\
type          &                    & performance & full-stack & scale    \\
\hline
\multirow{2}{*}{DES}
              & gem5/Simics        & \xmark     & \cmark     & \xmark \\
              & ns-3/OMNeT++       & \cmark     & \xmark     & \cmark \\
\hline
Modular comp. & \makecell[l]{SimBricks/\\SplitSim} & * & \cmark & \cmark \\
\hline
\multirow{3}{*}{Live sim.}
              & Phantora           & \cmark     & $\dagger$  & \cmark \\
              & NEX                & \cmark     & \xmark     & \xmark \\
              & \textbf{\sys}      & \cmark     & \cmark     & \cmark \\
\hline
\multicolumn{5}{@{}l}{\scriptsize * Depends on component simulators. $\dagger$ Runs unmodified ML apps without an OS.}
\end{tabular*}
\vspace{-12pt}
\end{table}

\section{Simulation-Oriented Scheduling\budget{1.5}}
\label{sec:scheduling}

\sys coordinates live and modeled execution under shared simulated time (C1) in two steps.
First, it represents each coordinated execution thread as a \emph{vtask} and tracks its simulated progress using a common virtual time (\emph{vtime}) (\autoref{sec:scheduling:vtask}).
Second, it groups vtasks into \emph{domains} that bound their vtime skew while accommodating temporary inactivity and different synchronization requirements (\autoref{sec:scheduling:dispatch}).


\subsection{Unified Virtual Time Accounting}
\label{sec:scheduling:vtask}

\para{Vtask abstracion.}
A vtask is a user-space thread whose execution participates in simulated-time coordination.
\sys utilizes this abstraction to decouple vtime accounting from the dispatch logic.
Each vtask exposes a vtime to the scheduler,
while an execution-specific accounting backend determines how that vtime advances. 
This separation lets the scheduler coordinate heterogeneous execution uniformly.

\sys supports two accounting backends:

\para{Clock-derived vtime.}
A \emph{live vtask}, such as a KVM vCPU thread,
executes the simulated component directly on hardware and derives vtime from its hardware execution time.
A configurable multiplier can scale this mapping when the target component runs at a different speed from the simulation host.
For a live vCPU, \sys implements clock-derived accounting by adapting KVM's paravirtual clock (pvclock)~\cite{pvclock}.
Stock KVM advances the guest clock with host wall time even when the vCPU is preempted and reports the preempted periods as steal time;
\sys instead adjusts the TSC (Time Stamp Counter)~\cite{intel64} offset at guest entry so that the clock advances only during guest execution.
While the vCPU runs, \sys extrapolates its vtime from the same clock state.
At guest exit, \sys folds the executed guest time into the vtask's vtime.
Therefore, the guest clock and the scheduler observe the same simulated progress.

The same clock-derived principle also applies to physical devices used as live components.
Rather than modeling their internal execution, \sys measure their physical execution time and translate it into simulated-time progress or delay.
For example, our live-DPU case study measures each packet's physical DPU processing time and schedules its completion after the corresponding vtime delay (\autoref{sec:evaluation:cases}).

\para{Model-driven vtime.}
A \emph{modeled vtask},
such as a QEMU device-model thread or ns-3 event-loop thread,
advances vtime according to its modeling logic.
They report accumulated vtime advances through a per-task \emph{run page} shared with the \sys kernel.
On the common path, the kernel consumes pending vtime advances asynchronously to avoid a system call for each report.
When a vtask must ensure that an advance has been consumed before continuing, 
it can instead use a synchronous \texttt{ioctl}.
Host-side execution itself does not advance a modeled vtask's vtime.
\sys uses a watchdog that identifies stalled vtasks and emits a warning for faulty components that stop making progress.

\subsection{Domain-Based Scheduling}
\label{sec:scheduling:dispatch}

\begin{figure}[t]
\centering
\includegraphics[width=\linewidth]{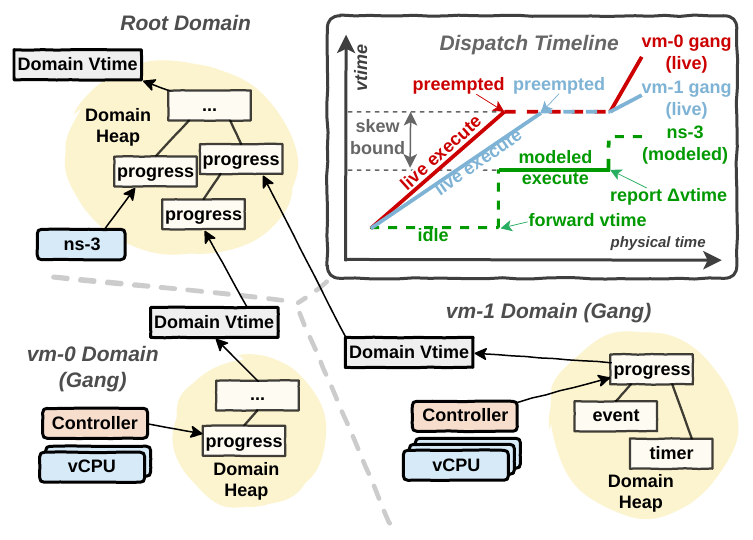}
\vspace{-14pt}
\caption{Domain-based scheduling for an example simulation with two live VMs and one modeled ns-3 component.
The domain hierarchy and constraints coordinate their progress, while the timeline illustrates skew-bounded dispatch across live and modeled execution.}
\vspace{-10pt}
\label{fig:scheduling}
\end{figure}

A common vtime makes the progress of heterogeneous vtasks comparable,
but coordinating that progress raises several additional challenges.
First, runnable vtasks should progress independently when possible to exploit parallelism,
while their vtime divergence remains bounded.
Second, a vtask that cannot currently make progress should not unnecessarily constrain others.
Third, some synchronization policies cannot be enforced through normal scheduler hooks alone.
Finally, synchronization requirements may vary across vtasks.
\sys addresses these challenges through domains, which abstract a group of vtasks that must stay synchronized.
\autoref{fig:scheduling} illustrates these mechanisms using an example simulation with two live VMs and one ns-3 modeled
component.




\para{Skew-bounded dispatch.}
A domain coordinates its members through constraints on how far they may advance.
\sys represents each constraint as a \emph{constraint node} carrying a simulated-time limit.
The domain maintains these nodes in a min-heap, called the \emph{domain heap}.
The minimum defines the \emph{domain vtime}.

Each domain $D$ has a \emph{skew bound} $bound_D$, which limits how far its members may progress independently.
Analogous to lookahead in conservative PDES~\cite{pdes,unison,simbricks}, it provides a window for independent simulated-time progress.
Whereas PDES derives lookahead from lower bounds on future event timestamps, such as communication delays, \sys applies a configured bound uniformly across a domain.
As shown in \autoref{sec:evaluation:micro}, keeping the skew bound below the smallest interaction delay within the domain
is important for preserving the timing fidelity of modeled interactions.

The skew bound exposes an accuracy--performance trade-off: a larger bound allows greater independent progress and parallelism, whereas a smaller bound reduces timing error from vtime skew at the cost of more synchronization.
The same heap can hold other constraints besides member progress.
For example, \autoref{sec:ipc} later uses this mechanism to enforce event visibility.

\para{Handling inactive vtasks.}
A vtask becomes \emph{inactive} when it cannot progress until a future event, e.g., a halted vCPU or a device awaiting I/O. \sys then removes its progress node from the heap while preserving its vtime. For timed reactivation, an \emph{event node} records the target vtime; when due, \sys advances the vtask to that time and restores its progress node. For untimed reactivation, \sys advances it to the current domain vtime before reinsertion.

\para{Domain controllers.}
To support more flexible scheduling policies, 
\sys allows each domain to attach a \emph{domain controller}, 
a kernel thread that performs domain-wide coordination beyond ordinary scheduler hooks.
The controller can monitor running members, update their scheduling state, and
interrupt them when necessary.
For example, a live-vCPU domain uses a controller to enforce its skew bound
without relying on periodic scheduler ticks that would perturb guest execution.
The controller extrapolates the vtime of running vCPUs, updates their progress
nodes, and preempts any vCPU that becomes ineligible.

\para{Heterogenous synchronization.}
Different vtasks within the same domain may require different skew bounds.
Applying the tightest requirement to one flat domain would unnecessarily constrain all members,
so \sys supports nested domains.
Each child domain enforces its own synchronization policy
and participates in its parent as a single member represented by the child's domain vtime.

Tightly coupled live vtasks may require near-zero vtime skew.
Simply using a small skew bound can be inefficient
as even a small lead can repeatedly make a member ineligible.
\sys therefore provides a \emph{gang}, a specialized domain whose controller
co-schedules its members on separate physical CPUs~\cite{gang}.
Concurrent execution keeps their progress closely aligned, while the gang appears to its parent as one coordinated member.

An SMP live VM is the primary use case for a gang.
Its vCPUs share a machine clock and guest memory,
so divergent execution progress can expose inconsistent guest-visible timing and abnormal shared-memory behavior.
As shown in \autoref{fig:scheduling}, each live VM forms a gang within the larger simulation domain.
The same mechanism can coordinate other tightly coupled components, such as VMs sharing disaggregated memory.

\section{Live Memory Hierarchy Management}
\label{sec:memory}

\sys keeps memory functional live while modeling its performance (\textbf{C2}). 
It progressively uses three mechanisms to bridge the gap between host and target memory behavior.
First, \emph{cells} isolate each VM within a controlled host-resource substrate (\autoref{sec:memory:cell}). Second, \emph{profile-guided vtime scaling} maps memory-stall exposure from host to target timing (\autoref{sec:memory:scale}). Finally, \emph{memory-region attribution} assigns stalls to memory regions and scales each by its target characteristics (\autoref{sec:memory:attribute}).


\subsection{Host-Interference Control with Cells}
\label{sec:memory:cell}

A \emph{cell} is \sys's unit of host-resource isolation for a live VM.
A cell binds the host resources needed by one live VM.
It assigns the VM dedicated CPUs and memory capacity on a host NUMA node,
hile constraining the shared cache capacity and memory bandwidth available to it.
When QEMU attaches a VM, \sys binds its guest RAM to that node,
populates the mapping, and keeps it resident.

\sys realizes these controls using cpuset-style CPU and NUMA placement~\cite{cpusets},
memory-capacity admission, and resctrl-based cache and bandwidth allocation.
It uses Intel Cache Allocation Technology (CAT)~\cite{intel64} for cache capacity allocation,
and Memory Bandwidth Allocation (MBA)~\cite{mba} constraining memory bandwidth.
CPU/NUMA placement controls where the VM executes and where its memory resides, 
while CAT and MBA constrain the shared resources it consumes.
Together, these mechanisms prevent co-located work from unintentionally changing the memory behavior observed by the live VM.

\subsection{Profile-Guided Virtual-Time Scaling}
\label{sec:memory:scale}

Cells remove unintended host interference but not systematic performance differences between host and target memory.
For example, host DRAM may be faster than a disaggregated memory.
\sys accounts for this difference by scaling each vCPU's vtime according to the portion of its execution exposed to memory latency.

Consider an accounting window in which
a fraction $f$ of the vCPU's execution time is exposed memory stall,
and the remaining fraction $1-f$ is unaffected execution.
Given a target-to-host memory-latency ratio $r$, \sys computes the vtime scale $s=(1-f)+fr=1+f(r-1)$.
Thus, only the execution time exposed to memory latency is scaled;
the remaining execution is unchanged.

\sys estimates $f$ using performance-monitoring unit (PMU) counters.
On Intel processors, it tracks \path{CYCLE_ACTIVITY.STALLS_L3_MISS},
which counts cycles in which execution stalls while a demand load misses the last-level cache~\cite{intel64}.
This measures the memory delay exposed to execution while naturally excluding latency hidden by memory-level parallelism.
The counter is attached to each vCPU and sampled during guest execution.
\sys smooths the resulting scale with an exponentially weighted moving average
and contributes it to the vCPU's vtime multiplier.

\subsection{Memory-Region Attribution Backends}
\label{sec:memory:attribute}

When a guest accesses memory regions with different target-to-host latency ratios,
a single aggregate stall fraction is insufficient.
We use an emerging configuration with guest-local DRAM and a disaggregated-memory region~\cite{cxl} as a running example.
\sys defines a hardware-independent attribution interface
whose backends report per-region stall shares to the common vtime scaler.
\sys currently provides three such backends.

\para{NUMA-source backend.}
On multi-NUMA Intel hosts, \sys places guest-local memory on the vCPU's NUMA node and a disaggregated-memory region on a remote NUMA node.
PMU events that distinguish local and remote DRAM accesses~\cite{intel64} allow \sys to estimate the stall share of each region.
Because the remote region already incurs additional host NUMA latency,
\sys accounts for this cost when mapping its stalls to the target latency.
This backend requires suitable NUMA topology and supports one disaggregated-memory class.

\para{Guest-assisted backend.}
When NUMA placement cannot distinguish memory regions,
\sys performs PEBS attribution inside the guest using a guest agent (\autoref{sec:multihost:dataplane}),
where sampled virtual addresses identify the accessed region.
The guest latency-weights the samples~\cite{pebs-precision}
and reports the resulting stall shares to the host,
which applies the corresponding target-to-host ratios.

\para{Physical-address backend.}
On AMD hosts, IBS Op samples memory-access latency together with,
when available, the accessed physical address~\cite{amd64}.
\sys maps sampled addresses to registered memory regions
and derives their latency-weighted stall shares directly,
without requiring NUMA placement or guest-side attribution.

\section{Simulation-Aware IPC\budget{1}}
\label{sec:ipc}


\sys provides simulation-aware IPC so that cross-component communication follows simulated rather than host timing (C3). We now introduce its three fundamental components.

\begin{figure}[!t]
\centering
\includegraphics[width=0.8\linewidth]{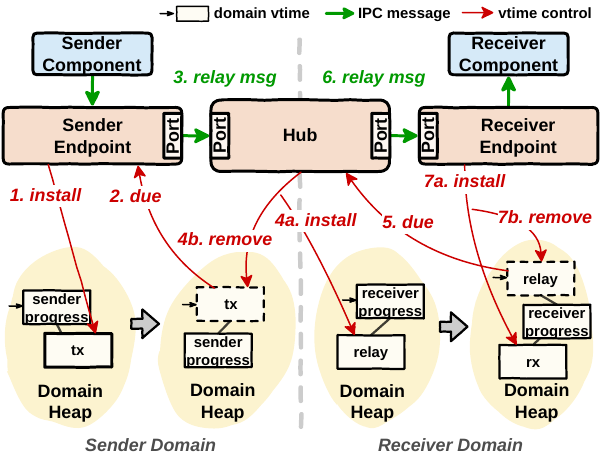}
\vspace{-6pt}
\caption{Message path and visibility-time enforcement.}
\vspace{-12pt}
\label{fig:ipc-path}
\end{figure}

\subsection{IPC Abstractions}
\label{sec:ipc:model}

As shown in \autoref{fig:ipc-path},
\sys uses \emph{hubs} and \emph{endpoints} to compose components and communication fabrics.

\para{Hubs.}
A hub models a communication fabric such as Ethernet, routing messages among attached endpoints and capturing the performance effects of traversing the fabric.
The hub operates on variants of \sys messages, which are based on a common abstraction across different fabric protocols.

\para{Endpoints.}
An endpoint abstracts a component's communication interface towards the hub, translating between the interface-specific I/O and variants of \sys messages.
Further, it can model the performance effects of traversing an interface such as a NIC.
Endpoints attach via the native API different components already expose for integrating with some communication fabric.
For an Ethernet interface, for example, a live VM exposes a QEMU vhost-user device as the API,
whereas a modeled component such as ns-3 exposes file-descriptor net devices.

\para{Ports.}
Hubs and endpoints have ports, which connect pair-wise to exchange the \sys messages. 
Ports provide only this relay interface and perform neither routing nor performance modeling and are therefore oblivious to the \sys message message variant being used.

\subsection{Enforcing Visibility-Time Delivery}
\label{sec:ipc:visibility}

To synchronize communication in simulated time,
\sys attaches a \emph{visibility vtime} to each message, which indicates when the next node in the communication path may observe it.
\sys enforces these times using the domain vtime heaps from
\autoref{sec:scheduling}.

Each endpoint is associated with its component's domain.
When an endpoint-local operation, such as NIC transmission or reception,
incurs modeled delay,
the endpoint buffers the message and inserts an event node at its visibility vtime.
In \autoref{fig:ipc-path}, the sender endpoint first installs the \emph{tx} node (1).
When the node becomes due (2), the endpoint relays the messaged to the hub (3).

Each relay establishes the next vtime constraint in its destination domain before the constraint in the source domain to ensure that an in-flight message always has a vtime constraint in some domain, preventing simulation progress from overtaking the visbility vtime.
After routing the message and applying the modeled fabric delay,
the hub therefore installs an event node at the resulting visibility vtime in the receiver domain
\emph{before} releasing the sender-side constraint (4a--4b).

\subsection{Modeling the Communication Fabric}
\label{sec:ipc:engines}

Communication fabrics vary widely in modeling complexity:
some require only parameterized link behavior,
while others require programmable forwarding or a detailed external simulator.
The hub abstraction lets \sys support these choices without changing
the endpoint and port interfaces.
\sys supports three types of fabric implementation:
built-in kernel models, programmable eBPF models, and external modeled components.

\para{Built-in models.}
Built-in hubs implement forwarding and timing entirely in the kernel
and expose configurable performance parameters.
For Ethernet, \sys provides a point-to-point link and a switch.
They suit common fabrics whose behavior is fixed while performance parameters vary.

\begin{figure*}[!t]
\centering
\includegraphics[width=\textwidth]{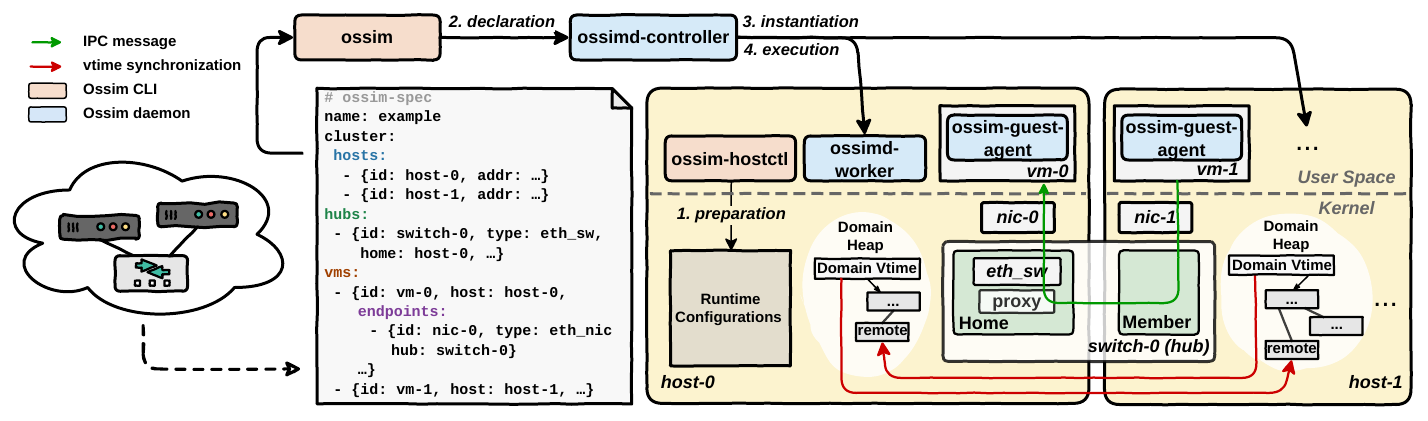}
\vspace{-20pt}
\caption{Running one simulation on a \sys cluster.
The control plane carries the simulation from preparation to execution, after which the worker kernels exchange frontier updates and IPC messages directly.}
\vspace{-7pt}
\label{fig:multihost}
\end{figure*}

\para{Programmable models.}
For customized fabrics, \sys uses a per-message eBPF hook for hub forwarding and timing while keeping the message path in the kernel. The hook selects the destination port and updates visibility vtime, with eBPF maps storing topology, routing, and contention state. \sys provides \texttt{eth\_ebpf} for this purpose.

\para{Modeled fabrics.}
More complex fabrics can run as external modeled components
with their own state and event loops.
They connect through the same endpoint--hub interface,
while \sys enforces visibility-time delivery at the boundary.
For example, an ns-3 topology can run as a modeled component
connected to live-VM endpoints through point-to-point link hubs,
while retaining ns-3's internal scheduler and network models.

\section{Multi-Host Coordination\budget{1}}
\label{sec:multihost}

\sys scales beyond one physical machine by coordinating hosts into a \sys cluster (\textbf{C4}).
\sys separates multi-host coordination into two parts:
a kernel \emph{data plane} preserves simulation semantics across host boundaries during execution (\autoref{sec:multihost:dataplane}), 
while a user-space \emph{control plane} configures the cluster and manages the lifecycle of a simulation (\autoref{sec:multihost:controlplane}).

\subsection{The Data Plane}
\label{sec:multihost:dataplane}

The data plane extends \sys's local semantics across host boundaries.
Though memory isolation remains host-local, the scheduling and IPC guarantees of \sys must be preserved with additional mechanisms to span across participating hosts.
\sys preserves these guarantees by having host kernels exchange vtime progress and timed IPC messages directly through inter-host channels.

\para{Cross-host vtime synchronization.}
Cross-host vtime synchronization preserves bounded-skew execution across hosts.
Each host publishes its root domain's vtime to its peers, which represent the received value with a \emph{remote constraint node} in their local domain heap.
This node constrains domain progress under the same bounded-skew dispatch rule in \autoref{sec:scheduling:dispatch}.
A host sends an update when its root-domain vtime advances sufficiently. To deal with the very uncommon case of lost updates, a host periodically retransmits its latest vtime.

\para{Cross-host IPC.}
Cross-host IPC extends a logical hub across hosts without changing its interface or timing semantics.
Endpoints attach to the same hub regardless of where their peers reside,
while the hub engine continues to determine routing and message visibility times as in \autoref{sec:ipc}.
Thus, distribution remains transparent to both endpoints and hub engines.

A cross-host hub represents remote endpoints through \emph{proxy endpoints}.
\sys realizes one logical hub with a \emph{home} instance and \emph{member} instances on remote hosts.
At the home instance, each remote endpoint is represented by a proxy endpoint,
allowing the hub engine to process it exactly like a local one.
\sys then transports the message and its visibility metadata to the corresponding member using reliable kernel TCP.
This preserves modeled loss semantics while allowing the same distribution mechanism 
to compose with different hub engines without distributed variants of their modeling logic.

\subsection{The Control Plane}
\label{sec:multihost:controlplane}

The control plane turns a cluster-wise simulation description into a running experiment.
Each host runs a worker daemon (\sysdw) that owns its local simulation resources,
and one controller daemon (\sysdc) coordinates operations across workers.
Every \sys host runs as a worker that owns its local simulation resources, while a controller coordinates operations across them.
A client acccesses the \sys cluster through the controller.
As \autoref{fig:multihost} shows, a simulation proceeds through preparation, declaration, instantiation, and execution.

\para{Preparation.}
Preparation establishes a known measurement state on each host.
\syshostctl administers each host before it participates in a simulation,
configuring and verifying measurement-relevant settings such as CPU power management, idle states, and reserved CPUs.
For machine-dependent parameters, \syshostctl derives appropriate values using a bundled host profiler.

\para{Declaration.}
Declaration specifies both the target simulation and its mapping onto the physical cluster.
A YAML specification submitted through \syscmd first describes the logical simulation: its components, their configurations, IPC hubs, and connections.
It maps these objects onto physical resources by naming participating workers and specifying component placement and resource assignments.
\sysdc resolves the specification into a cluster-wide plan and validates its references, placement, and resource-sharing constraints before sending 
to the workers.

\para{Instantiation.}
Instantiation creates the resources required by the validated simulation specification.
The controller sequences cluster-wide operations, while each worker creates and owns the component processes, kernel objects, and other backing resources that are assigned.
A prepare/commit barrier completes instantiation only after every worker reports that its local resources are ready.

\para{Execution.}
Execution runs the instantiated system and controls workloads inside its components.
To avoid paying synchronization overhead during guest boot and experiment setup, 
\sys can initially run components without vtime synchronization 
and switch to synchronized execution before entering the target region of interest.
The optional in-guest \sysagent communicates over vsock and supports features such as command execution and result collection.
It also carries the guest-assisted memory-attribution reports described in \autoref{sec:memory}.
For live VMs without \sysagent, \sys provides a console interface as a fallback for interactive guest control.

\section{Evaluation\budget{2.5}}
\label{sec:evaluation}

We evaluate whether \sys combines full-stack fidelity with practical simulation performance.
We first measure end-to-end accuracy and speed (\autoref{sec:evaluation:e2e}),
then isolate the mechanisms and their accuracy--performance tradeoffs (\autoref{sec:evaluation:micro}).
Finally, two case studies examine composition with different network models
and live hardware (\autoref{sec:evaluation:cases}).

\subsection{Implementation and Experimental Setup}
\label{sec:evaluation:setup}

\para{Implementation.}
\sys adds 42\,K lines to Linux, 3.0\,K to QEMU, 2.3\,K to ns-3, and
13.7\,K in \syslib for control-plane and operator support, plus roughly
15\,K lines of build, operator, and workload tooling.
Coding agents assisted implementation and debugging.

\para{Testbed and compared setups.}
Experiments use three identical hosts, each with dual Intel Xeon Gold 6326
processors (64 CPUs across two NUMA nodes) and a Mellanox ConnectX-5 NIC,
connected through a switch with 10\,Gb/s links.
We compare \sys with \emph{physical}, the physical ground truth;
\emph{stock KVM}, unmodified KVM/QEMU with Linux-bridge networking and
\texttt{netem} for latency or bandwidth shaping when needed; and
\emph{gem5 composition}, SimBricks~\cite{simbricks} with gem5 end hosts.
The hardware-backed setups use the same hosts, Ubuntu 24.04 LTS, and Linux 6.18;
stock KVM uses QEMU 8.2.
Target machines default to 4 CPUs and 16\,GB of DRAM.

A single pre-experiment run of \sys's host profiler calibrates machine-dependent
parameters (\autoref{sec:multihost:controlplane}).
All \sys experiments reuse this profile, which is independent of the workloads
and their target results.

\subsection{End-to-End Fidelity and Performance}
\label{sec:evaluation:e2e}

\begin{table}[t]
\centering
\caption{End-to-end fidelity and simulation performance.
$N$ is the number of server instances.}
\vspace{-4pt}
\label{tab:e2e-exps}
\scriptsize
\setlength{\tabcolsep}{1.5pt}
\renewcommand{\arraystretch}{1.1}
\begin{tabular*}{\columnwidth}{@{\extracolsep{\fill}}lclrrrr@{}}
\hline
Workload & $N$ & Metric & Physical & \shortstack{Stock KVM\\(err)}
& \shortstack{\sys\\(err)} & \shortstack{Wall /\\Vtime (s)} \\
\hline
memcached & 2 & p50 (\textmu s) & 66 & 127 (93.4\%) & 66 (0.0\%) & 155 / 60 \\
Hotel (K8s) & 2 & p50 (\textmu s) & 76 & 107 (40.2\%) & 81 (6.1\%) & 163 / 60 \\
Hotel (dist. K8s) & 3 & p50 (\textmu s) & 61 & 95 (54.9\%) & 58 (5.4\%) & 171 / 60 \\
\multirow{2}{*}{TPC-C (MySQL)}
& \multirow{2}{*}{2} & p50 (ms) & 7.74 & 13.01 (68.1\%) & 7.63 (1.5\%)
& \multirow{2}{*}{623 / 60} \\
& & tpmC & 38,706 & 24,118 (37.7\%) & 39,466 (2.0\%) & \\
\shortstack[l]{Redis\\\quad primary/backup}
& 3 & p50 (\textmu s) & 84 & 106 (25.3\%) & 87 (3.2\%) & 167 / 60 \\
TPC-DS 99 (Spark) & 3 & runtime (s) & 5.09 & 20.07 (294.4\%) & 4.63 (9.0\%) & 162 / 6.72 \\
\hline
\end{tabular*}
\vspace{-12pt}
\end{table}

\sys runs unmodified distributed stacks with end-to-end metrics within 9\% of
physical measurements.
\autoref{tab:e2e-exps} covers memcached, Kubernetes-based hotel reservation on
one or multiple hosts, MySQL TPC-C, Redis primary/backup, and a three-instance
Spark TPC-DS workload.
Across seven reported metrics, its median error is 3.2\%, whereas stock KVM
deviates by 25--294\%.
Software compatibility alone therefore does not preserve target performance;
simulation must also control execution, resources, and communication.

This fidelity remains practical to obtain: \sys requires 2.6--24 host seconds
per simulated second, whereas no corresponding gem5 end-to-end experiment
completes within the 24-hour timeout.

\subsection{Accuracy and Performance Microbenchmarks}
\label{sec:evaluation:micro}

\begin{figure*}[!t]
\centering
\includegraphics[width=\textwidth]{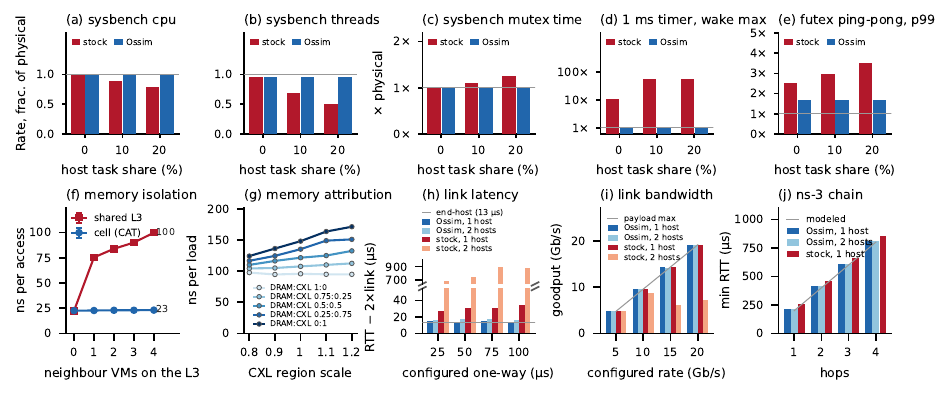}
\vspace{-20pt}
\caption{Mechanism-level accuracy. 
(a)--(e) evaluate scheduling under different host CPU background load;
(f)--(g) evaluate memory interference isolation and per-region performance scaling;
(h)--(j) evaluate modeled communication on one host and across hosts.
Thin gray lines indicate the optimal case in (a)--(e), (i), and (j), and the measured end-host delay in (h).}
\vspace{-7pt}
\label{fig:micro}
\end{figure*}

\begin{figure}[t]
\centering
\includegraphics[width=\linewidth]{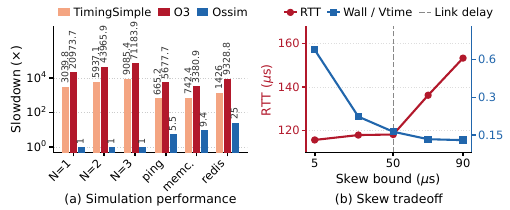}
\caption{(a) Simulation performance against gem5 composition.
(b) RTT and simulation cost (wall-clock seconds per simulated second) versus skew bound; the dashed line marks the modeled one-way latency.}
\label{fig:perf-skew}
\vspace{-12pt}
\end{figure}

To explain these end-to-end results,
we isolate scheduling, memory, and communication control (\autoref{fig:micro}), 
then examine simulation cost and skew-bound sensitivity (\autoref{fig:perf-skew}).
Accuracy is measured against matching physical hardware when available and configured models otherwise.
Because the gem5 CPU models are not calibrated to our processors, we use them
only for simulation-performance comparisons.

\para{Scheduling (\autoref{sec:scheduling}).} \sys's simulation-aware scheduling keeps guest execution and wake-up timing close to physical under host CPU contention.
We run \texttt{sysbench} CPU, threads, and mutex tests, a 1\,ms periodic timer,
and a two-thread futex ping-pong while background work consumes 0, 10, or 20\%
of each vCPU's physical core.

\autoref{fig:micro}(a)--(c) show the effect on guest execution.
Both stock KVM and \sys stay close to physical without background load.
As load increases, stock KVM suffers 
lower throughput or longer runtime.
\sys instead stays near its isolated result because guest vtime advances with accounted execution rather than host wall time.

For wake-up timing, \autoref{fig:micro}(d)--(e) shows the contrast appears even without background load.
Maximum timer wake-up latency is 10.4$\times$ physical under stock KVM but only 1.1$\times$ under \sys.
Background load further raises the stock-KVM timer latency to 56$\times$ physical and increases futex p99 latency,
whereas \sys's coordinated dispatch keeps both probes close to physical.

\para{Memory management (\autoref{sec:memory}).}
Beyond execution timing, \sys controls both memory interference and region-specific performance.
We first test isolation with a dependent-load pointer chase in one VM,
while up to four neighboring VMs stream through the last-level cache on other cores of
the same socket (\autoref{fig:micro}(f)).
Without isolation, latency rises from 22.5 to 100\,ns;
by enforcing cells, it remains 22.7--23.1\,ns, within 3\% of the isolated case.

Whereas isolation stabilizes the live memory substrate, per-region scaling
adjusts its modeled performance (\autoref{fig:micro}(g)).
A memory-intensive chase mixes local DRAM and CXL accesses while only the CXL
region's scale varies from 0.8 to 1.2.
All DRAM workloads remain stable, all CXL workloads track the scale,
and mixed workloads follow targets weighted by their CXL access shares.
This validates \sys's region-specific stall attribution and timing adjustment.

\para{IPC and multi-host coordination (\autoref{sec:ipc}, \autoref{sec:multihost}).}
Beyond controlling each live VM, \sys must preserve modeled communication
between components, including across hosts.
For \autoref{fig:micro}(h)--(i), stock KVM uses a Linux bridge with
\texttt{netem} and runs without background load.

\autoref{fig:micro}(h) first evaluates modeled latency.
We vary one-way link latency from 25 to 100\,$\mu$s and report $\mathrm{RTT}-2\times\mathrm{link\ latency}$,
which should remain near the measured 13\,$\mu$s end-host contribution.
\sys leaves only 14--15\,$\mu$s on one host and 16--17\,$\mu$s across hosts,
whereas stock KVM leaves 27--35\,$\mu$s and 693--892\,$\mu$s, respectively.
This gap arises since \texttt{netem} adds modeled delay atop the
physical path, whereas \sys controls message visibility in simulated time.

This decoupling from the physical path also
lets \sys reproduce modeled bandwidth beyond the physical 10\,Gb/s link (\autoref{fig:micro}(i)).
Both setups match the configured bandwidth on one host,
but across hosts stock KVM is capped by the physical link.
In contrast, for modeled 15 and 20\,Gb/s links, \sys delivers 14.20 and
19.10\,Gb/s, within 1.1\% of the respective payload maxima of
14.35 and 19.13\,Gb/s.

Finally, \autoref{fig:micro}(j) shows that the same timing control extends to an external ns-3 network model.
For a chain adding 100\,$\mu$s one-way latency per hop,
both systems reproduce the expected 200\,$\mu$s RTT increase per hop.
However, \sys remains only 12--14\,$\mu$s above the model, versus 57--60\,$\mu$s for stock KVM with ns-3's TAP integration.
Distributing \sys's VMs across two hosts, with ns-3 on one of them,
adds only about 1\,$\mu$s while preserving the same slope.
Our stock-KVM baseline has no corresponding two-host configuration because
its ns-3 TAP integration connects the live guests and ns-3 only within
one host.

\para{Simulation performance.}
Whereas the preceding experiments evaluate accuracy, \autoref{fig:perf-skew}(a)
examines the simulation-performance benefit of live end-host execution over gem5 composition.
For compute-intensive \texttt{sysbench}, increasing SMP width from one to three CPUs raises gem5's cost with both TimingSimple and O3 models.
In contrast, \sys's cost remains nearly flat
because additional vCPUs run directly on additional physical cores in parallel.

We therefore give every simulated host only one CPU in the distributed gem5 experiments,
which favors the gem5 composition.
Even with this configuration, \sys requires 5.5--25 wall-clock seconds per simulated second,
compared with 665--1426 for gem5 composition with SimpleTiming CPU and 3381--9329 for O3.
Across the corresponding workloads, \sys is 57--121$\times$ faster than
TimingSimple and 360--1032$\times$ faster than O3.
This shows that live execution removes the end-host simulation cost
that otherwise dominates a full-stack modular composition under typical workloads.

\para{Skew-bound tradeoff.}
Live execution avoids detailed CPU simulation,
but its coordination cost still depends on how tightly components synchronize.
\autoref{fig:perf-skew}(b) examines the accuracy--performance tradeoff
by sweeping the bound from 5 to 90\,$\mu$s for two VMs in separate gangs.
The VMs exchanging pings over a link with 50\,$\mu$s one-way latency.
Increasing the bound from 5 to 50\,$\mu$s keeps RTT near 118\,$\mu$s while
reducing cost from 0.715 to 0.160 host seconds per simulated second
(4.5$\times$).
Beyond the modeled one-way delay, however, RTT rises by 18 and then 35\,$\mu$s
for little reduction in cost.
These results suggest choosing a bound no larger than the smallest modeled
one-way delay relevant to the experiment.

\subsection{Case Studies}
\label{sec:evaluation:cases}

Having examined fidelity and performance,
we demonstrate the flexibility and generality of \sys with two case studies.
We first compose the same live VMs with different underlying network implementations,
and then add a physical DPU as a live component.

\para{Composing a fat-tree network.}
The first case study changes the network model without changing VM interfaces.
We connect sixteen one-vCPU VMs to a $k{=}4$ fat-tree with 10\,$\mu$s latency
per hop, using either an eBPF hub or external ns-3 component
(\autoref{sec:ipc:engines}).
The eBPF hub implements topology, routing, and timing in a programmable hub,
whereas ns-3 provides the network as an external modeled component which allows adding additional background traffic with modeled hosts.
As \autoref{tab:cases}(a) shows, both network implementations
reproduce the expected 40\,$\mu$s RTT increase between topology tiers.

\para{Composing a physical DPU.}
Whereas the fat-tree case varies a modeled fabric,
the second case extends live execution to a physical BlueField-2 DPU for packet processing offloading.
\sys provides an in-kernel backend that connects the physical DPU to a VM's NIC endpoint through two dedicated virtual functions (VFs).
Packets are submitted through one VF and the DPU's rewritten packets are received through the other.
For traffic that comes from the opposite direction, the VF roles reverse.
The backend measures each packet's DPU processing time and schedules completion according to it in vtime.

Using this integration, we compare NAT in a gateway VM's Linux stack, DPU software, and the DPU ASIC.
Three live VMs---client, gateway, and server---connect through an Ethernet
switch hub; the DPU paths bypass the gateway's network stack.
To compare sensitivity to gateway load, we place 1,024 nonmatching NAT rules before the matching rule
and stress the gateway with workers that run inside it, measuring 100 UDP echoes per configuration after warmup.
CPU-path RTT rises from 70.04 to 163.12\,$\mu$s with increasing load, while DPU
software and ASIC paths remain near 72 and 54\,$\mu$s
(\autoref{tab:cases}(b)).

\begin{table}[t]
\centering
\caption{Case studies.
(a) Round-trip time between VM pairs in the $k{=}4$ fat-tree.
(b) Round-trip time of NAT through the gateway VM's CPU and DPU paths.}
\vspace{-5pt}
\label{tab:cases}
\scriptsize
\setlength{\tabcolsep}{2.5pt}
\renewcommand{\arraystretch}{1.1}
\begin{minipage}[t]{0.42\columnwidth}
\centering
\textbf{(a) Fat-tree RTT (\textmu s)}\\[2pt]
\begin{tabular*}{\linewidth}{@{\extracolsep{\fill}}lrr@{}}
\hline
VM pair & eBPF & ns-3 \\
\hline
Same edge & 59.7 & 57.8 \\
Same pod  & 99.4 & 98.1 \\
Cross pod & 139.8 & 137.5 \\
\hline
\end{tabular*}
\end{minipage}
\hfill
\begin{minipage}[t]{0.54\columnwidth}
\centering
\textbf{(b) NAT RTT (\textmu s)}\\[2pt]
\begin{tabular*}{\linewidth}{@{\extracolsep{\fill}}crrr@{}}
\hline
Workers & CPU & DPU-SW & DPU-ASIC \\
\hline
0 & 70.04 & 72.71 & 54.89 \\
4 & 149.16 & 71.43 & 53.95 \\
8 & 163.12 & 72.02 & 54.29 \\
\hline
\end{tabular*}
\end{minipage}
\vspace{-10pt}
\end{table}

\section{Related Work and Discussion\budget{0.40}}
\label{sec:related}

\noindent\textbf{Simulation acceleration.}
Apart from live simulation, prior work accelerates simulation through system and runtime
optimizations~\cite{dons,unison,klonet,legosim,splitsim,ash}, learned performance
models~\cite{routenet,routenet-fermi,mimicnet,deepqueuenet,m3}, and specialized hardware such as
GPUs and FPGAs~\cite{cunetsim,nsx,gedes,gem,miniature,firesim}.

\para{Simulation for distributed stacks.}
Other work targets specific cluster workloads, particularly GPU training~\cite{simai,spme,phantora},
by simulating both distributed training strategies and the underlying hardware behavior.
In contrast, \sys focuses on unmodified full-stack distributed software and its interaction with
underlying hardware at cluster scale.

\para{Resource isolation for guests.}
Partitioning hypervisors~\cite{lpar,nohype,jailhouse,bao} statically bundle fixed host resources for
each guest and isolate them through strict non-sharing of cache and memory hardware.
In contrast, \sys dynamically allocates host resources as simulated guests come and go to preserve
simulation accuracy under controlled sharing.

\para{Discussion.}
Oversubscribing a cell would require \emph{temporal} sharing of the live memory hierarchy.
Even if the scheduler time-slices vCPUs, an incoming live VM can inherit cache and prefetcher
state from its predecessor.
Preserving fidelity would require reconditioning that state.
Temporal cache reconditioning remains outside this work, and \sys does not support safe cell
oversubscription.
Beyond KVM-based VMs, integrating other virtualization
technologies~\cite{nvidia-vgpu,xsched,cricket,gpreempt,krypton,gshare,nixie} could extend \sys
to additional live hardware such as GPUs.

\section{Conclusion\budget{0.10}}
\label{sec:conclusion}

\sys takes an OS-level approach to cluster-scale full-stack simulation of unmodified distributed
software stacks by integrating simulation-oriented mechanisms into the Linux virtualization stack.
\sys keeps production software live while OS mechanisms control simulated time, memory,
communication, and their distribution across hosts.
We see \sys as an initial step toward \emph{simulation-native OS support}, a broader direction
beyond cluster-scale full-stack simulation.

\balance
\bibliographystyle{ACM-Reference-Format}
\bibliography{references}

\end{document}